\documentstyle[aps,twocolumn,prl]{revtex}
\input epsf

\begin{document}

\twocolumn[\hsize\textwidth\columnwidth\hsize\csname @twocolumnfalse\endcsname

\title{Simple model for reverse buoyancy in a vibrated granular system}
\author{G. Guti\'errez$^{1,*}$, O. Pozo$^{1}$, L. I. Reyes$^{1}$, R. Paredes V.$^{2}$, J. F. Drake$^{3}$ and E. Ott$^{3}$}
\address{$^{1}$Departamento de F\'{\i }sica, Universidad Sim\'{o}n Bol\'{\i }var,
Apartado 89000, Caracas 1080-A, Venezuela}
\address{$^{2}$Centro de F\'{\i}sica, Instituto Venezolano de Investigaciones
Cient\'{\i }ficas, Apartado Postal 21827, Caracas 1020-A, Venezuela}
\address{$^{3}$University of Maryland, College Park, Maryland 20742-2431}

\date{\today }
\maketitle
\widetext
\begin{abstract}
\noindent Large objects, immersed in a homogeneous granular system, migrate when subjected to vibrations. 
Under certain conditions large heavy objects rise and similar light ones sink to the bottom. 
This is called reverse buoyancy. We report an experimental study of this singular behavior, 
for a large sphere immersed in a deep granular bed. A simple mechanism is proposed to describe 
the motion of a sphere, inside a vertically vibrated granular system. When reverse buoyancy is observed, 
the measured vertical velocity of the immersed object, as a function of its density, shows a simple behavior. 
With a one-dimensional mechanical model that takes into account a buoyancy force and the frictional drag, 
we obtain the rising velocity for heavy objects and the sinking rate for light ones. The model yields a 
very good qualitative and quantitative agreement with the experiment. 
\end{abstract}


\preprint{HEP/123-qed}]

\narrowtext
\footnotetext{* Corresponding author, e-mail address: gustav@usb.ve}

Granular matter is presently a subject of extensive research, due to the interesting behavior frequently encountered in this kind of systems (dilatancy, segregation, arching, clustering, etc.), and the many applications in industry in areas like agriculture, food processing, geophysics, 
pharmacology, etc., that can profit from those investigations \cite{Jaeger,duranbook}.

Dry granular materials are difficult to mix homogeneously. Industrial processing involving mixtures of different grains very often lead to undesirable segregation of these. Properties of the grains like size, shape, density, etc., determine their separation in unexpected ways. It has become clear that segregation of grains under shaking, is a very rich but poorly understood phenomenon. Size segregation has caught the attention of many scientists leading to an extensive research on different aspects of the problem. This research has revealed the complexity involved and the many aspects of grain separation processes that are still unsolved.

Shinbrot and Muzzio \cite{shimbrot} reported that large heavy objects, called 
intruders, immersed in a granular medium, rise when shaken, while similar objects that are light migrate to the bottom. This novel effect, that shows the complexity and richness of the behavior of granular matter subjected to vibrations, was called reverse buoyancy. The above authors emphasized the role of inertia in the upward motion of the intruders, but the cause of the downward motion, for light objects, has gone unexplained. It has become clear that much remains to be learned from this intriguing behavior.

In the migration of a large object immersed in a deep granular bed there may be many factors playing a role simultaneously. Some of the factors are: small particles migrating to the void left out by a large object \cite{rosato,jullien},
convection currents \cite{knight,duran1,awazu,poschel}, arching 
\cite{duran2}, buoyancy force \cite{shisho}, and condensation \cite{hong}. We expect to find different regimes where one or more of these factors are dominant, depending on the way the system is shaken and on several characteristics of the granular medium or the intruder. 
It should be interesting to be able to separate different regimes and determine 
the relative weight of the different factors affecting the migration of large objects in a bed of small particles. This could increase significantly our understanding of granular matter and would certainly have a significant impact in industry. 

M\"obius {\it et al} \cite{mobius} have pointed out that the rising time of the intruder may be used as a sensitive probe of these different interactions. They also mention that increasing evidence indicates that differences in particle density affect size separation in granular mixtures. In this work, we have measured the velocity of migration for large objects with equal shapes, similar sizes and different densities, when reverse buoyancy is observed. With a very simple mechanical model we are able to determine the velocity of the intruder as a function of its density.

For our experimental set up we used spherical glass beads with diameters between 
$210$ and $250\mu m$, the intruders were spheres of approximately $2.5 cm$ in diameter. 
Hollow plastic spheres were filled with different materials to vary the density. 
The vertical vibrations were produced with a VTS100 electromagnetic shaker coupled to a Plexiglas rectangular container with volume equal to $15 \times 8\times 2.7 cm^3$. 
The frequency used was $f =11.7Hz$, the amplitude was fixed at $A=8.5mm$, so that the 
adimensional amplitude of the acceleration was $\Gamma= A (2\pi f)^2/g=4.7$, where $g$ is the acceleration of gravity. $\Gamma$  was monitored using an ADXL150 accelerometer. 
The container was filled with the glass beads up to a height of $11cm$ 
and the intruder was introduced in the medium at a certain depth of the container. Since the large spheres could be seen through the Plexiglas wall, a digital camera was used 
to register the position of the intruder as a function of time (see figure 1). The density of the intruder was varied from $0.26gr/cm^3$  to $7.7gr/cm^3$. These correspond to a range of approximately $20\%$ to $500\%$ of the density of the medium. This density was measured by sampling different volumes of the granular medium and measuring the weight with an electronic balance.

In figure 2 we show a typical graph of the vertical position of the intruder
 as a function of time for a heavy (a) and a light (b) intruder. The reference frame is located on the container with the positive y-axis in the upward direction. Figure 3 shows the velocity of the  intruders as a function of the ratio $\rho_i/\rho_m$, where $\rho_i$ is the density of the intruder and $\rho_m$ is the density of the granular medium. 
We can see that, for the measured densities, the average velocity of the heavy intruders 
is positive, and its magnitude increases monotonically with its density. 
While, for the light intruders, the average velocity is negative and its magnitude decreases 
with increasing density.

	In an attempt to explain the above results we formulate the following simple model: 
It is assumed that the medium fluidizes only in part of the cycle of oscillation. This was proposed by de Gennes in relation to an instability in a sand heap \cite{evesque}. 
He proposed a series of passive and active regimes. A passive regime occurs in each oscillation, when the grains are compacted and can be considered as a solid, and an active 
regime develops in the same cycle, when the granular system is fluidized. According to this approach, during the active fraction of the oscillation, the system is subjected to an apparent gravity, reversed upward. This reversal of the apparent gravity increases 
the relative interparticle motion of the medium causing a fluidization. 
This assumption was used in connection with instabilities observed in shaken sand 
in a U-tube \cite{drake}, a similar condition was used to describe a transition from 
a condensed state to a fluidized state in a column of beads \cite{duran3}, and to derive a criterion for the onset of covection \cite{shimbrot2}. When the sand is not fluidized it behaves like a solid and the intruder is trapped and does not move relative to the granular bed. When it is fluidized we assume that three forces are dominant: the weight, the drag and the buoyancy. In the noninertial reference frame of the container, during the active regime, 
we obtain the following equation of motion for the position $y(t)$ of the intruder:
\begin{equation}
\frac{d^2y}{dt^2}=g\left(\Gamma \cos\omega t-1\right)\frac{\Delta\rho}{\rho_i}-
\nu_0\frac{\rho_m}{\rho_i}\frac{dy}{dt} , \label{yseg}
\end{equation}
where $\omega=2\pi f$, $\Delta\rho=\rho_i-\rho_m$, and $\nu_0$ is a positive constant.

We have assumed that the drag force is proportional to the density of the granular medium and to the velocity of the intruder. The constant $\nu_0$ may depend on the size and the shape of the intruder, and the effective viscosity of the medium while 
fluidized. In this model,  the sand fluidizes when the term inside the parentheses, 
in equation (\ref{yseg}), is positive. This active regime occurs during the interval of time $\tau$ when the acceleration, due to the moving noninertial reference frame, points upward and its magnitude is greater than gravity. As we mentioned before, this condition can be seen as an apparent gravity pointing upward, causing an increase in the relative interparticle motion. The time interval $\tau$, in one cycle of the oscillation, goes from the instant the granular system becomes active to the time 
when the passive regime starts, and is given by 
 
\begin{equation}
\tau=\frac{2}{\omega}\cos^{-1}\left( \frac{1}{\Gamma} \right) .
\end{equation}

At the beginning of this time interval $\tau$ the velocity of the intruder relative to the noninertial 
reference frame is taken to be zero.
 
Integrating twice equation (\ref{yseg}) we calculate the displacement $\Delta y_\tau$ per cycle, during the time interval $\tau$. If we sum all the contributions per cycle in a time $t$ much greater than the period $T$ of the oscillations, the average vertical displacement $y(t)$ of the intruder is obtained,
\begin{equation}
y(t) = \left[ \frac{g}{\pi\nu_0}(tan\alpha -\alpha)\frac{\Delta\rho}{\rho_m} \right] t,
\end{equation}
where $\alpha=\omega\tau/2$, and we have neglected a transient term, assuming that $\nu_0(\rho_i/\rho_m)\gg\omega$.
The expression inside the square brackets is the average vertical velocity of the intruder.
This velocity can be expressed as,
\begin{equation}
v(t)= F(\alpha,\nu_0)\frac{\rho_i}{\rho_m}- F(\alpha,\nu_0) ,
\label{4}
\end{equation}
where,
\begin{equation}
F(\alpha,\nu_0)=\frac{g}{\pi\nu_0}(tan\alpha-\alpha) .
\end{equation}

The model predicts the following characteristics of the migration of a large 
object in a granular medium when this fluidizes in part of the cycle:
\begin{enumerate}
\item Heavy objects ($\rho_i>\rho_m$) rise at constant velocity.
\item Light objects ($\rho_i<\rho_m$) sink at constant velocity.
\item The dependence of the velocity on the ratio $\rho_i/\rho_m$ is linear
and the slope is equal to the negative of the intercept, on the velocity axis.
\end{enumerate}

Near $\Delta\rho=0$ other contributions may become dominant, but it is assumed, 
for simplicity, that the model remains valid near $\rho_m=\rho_i$.

In figure 3 we can see that the experimental results agree with the predictions of the model. The slope is equal to the magnitude of the intercept, within two significant figures. 
If $F(\alpha,\nu_0)=12mm/s$, we get that the value for the only free parameter is $\nu_0 = 840s^{-1}$. In the inset a plot of the velocity versus the radius of three large heavy spheres is shown. These measurements were made with three steel balls with different diameters, and the same density. This  result suggests that the velocity $v$ of the intruders depend linearly on the radius $r$ of the spheres. This is consistent with a linear dependence of the drag force on the transversal area of the intruders, which implies that the retarding force is proportional to $r^2\rho_m v$. Further measurements are needed to determine the dependence of the velocity on the size and shape of the intruder.

No points are reported very close to $\Delta\rho=0$. The reason is that, in such range, the velocity is very small and the motion becomes erratic. The same object can stay in one place for a long time and then drift slowly either upward or downward.

Our results are significantly different to those reported by M\"obius {\it et al} \cite{mobius}. In their experiment all the intruders rise, even if the density is less than of the granular medium. We believe that the reason for this is that they used tapping instead of sinusoidal oscillations to shake the container. We tested this hypothesis by shaking a container with sand, and using a light intruder. When the motion was approximately sinusoidal, the intruder would sink, and when we applied brief and distinct shocks the same intruder would rise to the top. A  distinct shock, seen from the reference frame on the container, produces a large acceleration downward, and then a smaller acceleration upward as the system relaxes back to its original position. This does not produce an effective reversal of the apparent gravity, during a sufficiently long time $\tau$. The mechanism involved in the migration of an intruder, may be different from the one found in a vibrated granular matter, when tapping is used.

We are presently studying in detail, the conditions for which reverse buoyancy can be observed. These conditions should give clues about the detailed mechanism of the cyclic fluidization and condensation of the granular medium subjected to vibrations.

Although the assumptions are crude, the predictions based on them work very well for our experiment. We see from the experimental results that density ratio can be a crucial parameter to describe the migration of large objects immersed in a medium of smaller particles. An important advantage of our model is that it gives a very simple explanation for the rising and sinking of large objects in vibrated granular beds,  in terms of the density ratio, for the case of reverse buoyancy. This experiment gives strong support to the idea that the oscillations can produce a series of active and passive regimes, where the granular medium can be considered a fluid, in part of the period of oscillation, and a solid in the remaining part. 

In conclusion we have provided a useful way of interpreting the physical origin of reverse buoyancy through measurements of the velocity of a large object immersed in a vibrated granular bed, and a simple model that indicates the role of a buoyancy force and the drag force. With this model we are able to predict the velocity of the intruder as a function of its density. According to the proposed model the observed behavior is a consequence of the fluidization of the granular medium in only part of the cycle.  A detailed understanding of how the granular medium is fluidized and the relative influence of various parameters is not yet available. It would be interesting to investigate the influence of ambient gas, and the effect of changing the physical properties of the grains. In particular, the presence of ambient gas may be necessary for the fluid-like behavior during the active phase. If this is so, then reverse buoyancy might be expected to be absent in a sufficiently hard vacuum. Indeed, in some situations, hard vacuum has been observed to result in an analogous suppression of heaping of vertically vibrated granular media\cite{behringer}.

GG would like to express special thanks to Chris Lobb for initiating him in the subject of granular matter, to the CSR and to the Nonlinear Dynamics Laboratory at the University of Maryland for the support and hospitality during that initial period. We are indebted to Sergio D\'{\i}az, for providing the shaker. Thanks to Hans Herrmann , Leonardo Trujillo, Alfred Cawthorne and Adolfo Rodr\'{\i}guez, for the useful discussions. This research was supported in part by FONACIT, under Grant S1-2000000624, and by the DID, of the Universidad Sim\'on Bol\'{\i}var.



\begin{figure}[tbp]
\caption{  Sequence of pictures from a digital video of the vibrating granular system, with an intruder. 
The top three images are of an intruder of low density $\rho_i = 0.26gr/cm^3$, from lefth to right  $t=0s$, $t=3.83s$ and $t=5.27s$.
The bottom three pictures are of an intruder of high density $\rho_i =7.7 gr/cm^3$, from left to right  $t=0s$, $t=0.77s$ and $t=1.10s$. 
The rectangles indicate the position of the intruder.}
\end{figure}

\begin{figure}[tbp]
\caption{ The measured vertical displacement for (a) heavy intruders and 
(b) light objects, as a function of time. The linear fit shows the agreement between the model and the experiment.}
\end{figure}

\begin{figure}[tbp]
\caption{ Vertical velocity of the intruder as a function of the density ratio, $\rho_i/\rho_m$. The solid line
represents a linear fit. The equation resulting from the fit shows the complete agreement beteen theory and experiment (see Eq.\ref{4}). 
The inset shows the dependence of the velocity of a heavy intruder on the radius of the sphere, 
for three steel balls of the same density. }
\end{figure}

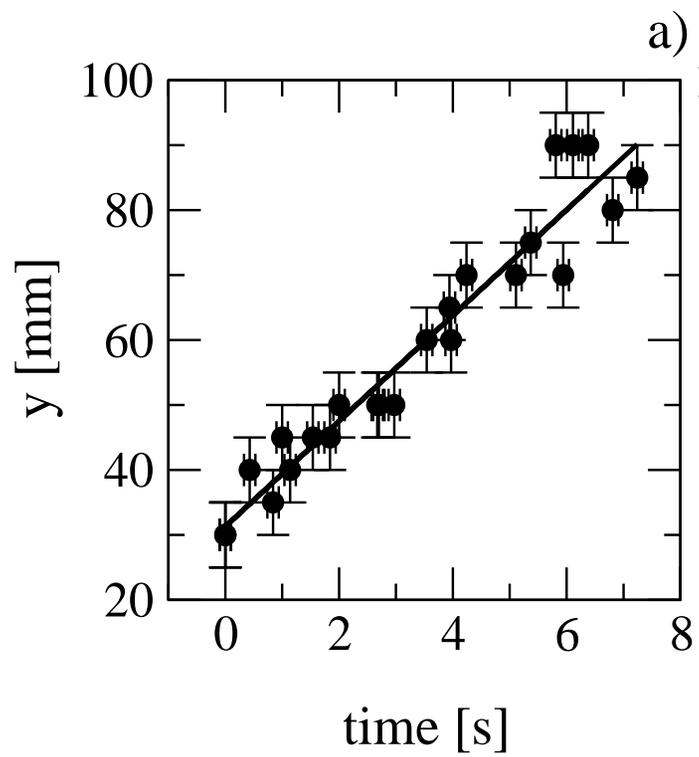
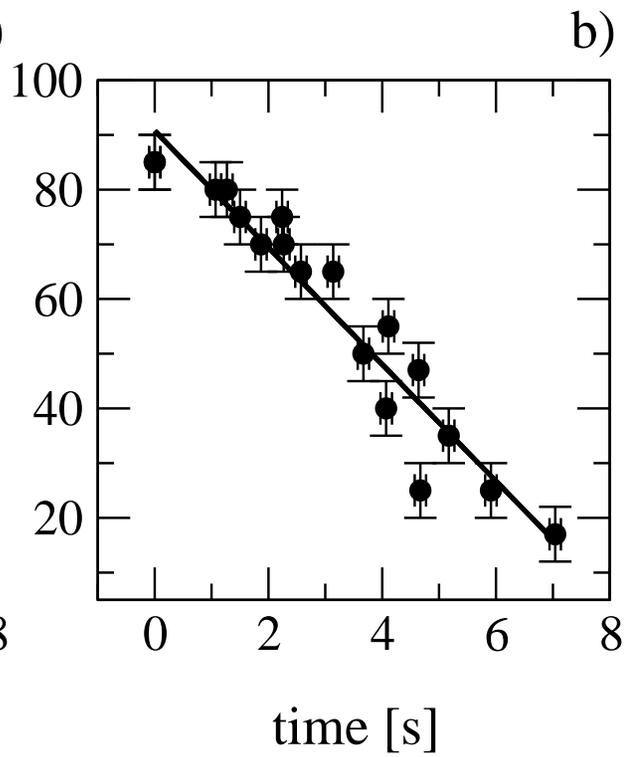
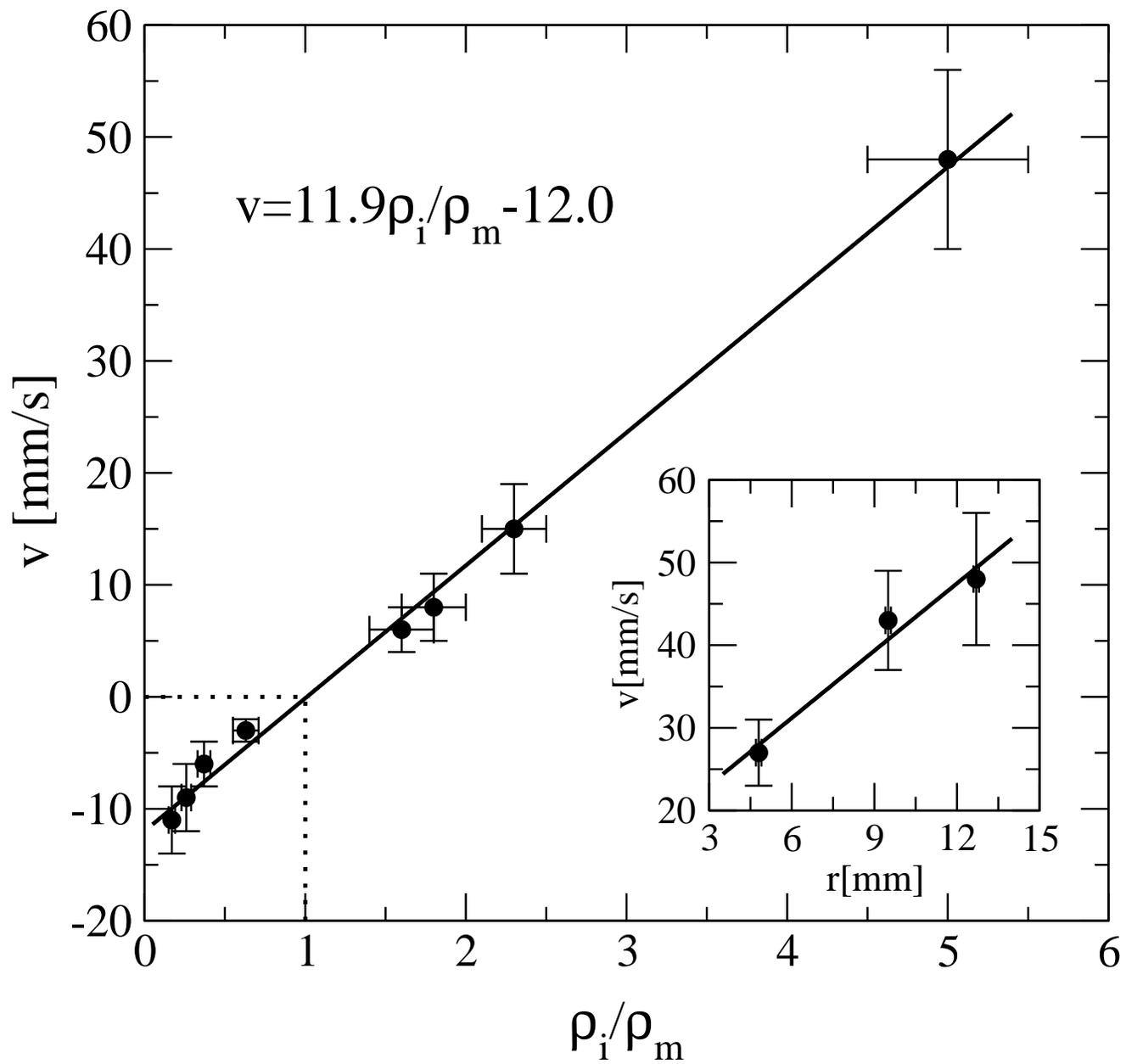
\end{document}